\documentclass[lettersize,journal]{IEEEtran}
\usepackage[T1]{fontenc}

\usepackage{amsmath,amsfonts}
\usepackage{algorithmic}
\usepackage{algorithm}
\usepackage{array}
\usepackage[caption=false,font=normalsize,labelfont=sf,textfont=sf]{subfig}
\usepackage{textcomp}
\usepackage{stfloats}
\usepackage{url}
\usepackage{verbatim}
\usepackage{graphicx}
\usepackage{cite}
\usepackage{booktabs}   
\usepackage{multirow}

\begin{document}
\pagestyle{plain}

\title{Iterative Self-Learning for Expressive Text-to-Speech Synthesis}

\author{Nicholas Sanders,
        Gustav Eje Henter,
        Simon King,
        Korin Richmond
\thanks{This work was partially supported by the UKRI CDT in NLP,
funded by the UKRI (grant EP/S022481/1), the University of
Edinburgh, Huawei, and the Wallenberg AI, Autonomous Systems and
Software Program (WASP) funded by the Knut and Alice Wallenberg Foundation.}}%

\maketitle
\thispagestyle{plain}
\pagestyle{plain}
\begin{abstract}
Expressive text-to-speech (TTS) systems that use explicit conditioning labels provide direct and interpretable control over expressive attributes, in contrast to reference-based or prompting-based approaches, but require labeled data. Obtaining these labels at scale is costly and time-consuming, yet no prior semi-supervised framework addresses this specific bottleneck. Existing semi-supervised TTS methods instead target scarcity of paired speech-text data or transcriptions. To address the scarcity of expressive labels, we propose an Iterative Self-Learning (ISL) framework for expressive TTS, built on Invert-Classify, a classifier-free method that recovers discrete expressive labels by inverting a frozen generative model. The framework iteratively pseudo-labels unlabeled speech using the current model, retrains on the combined labeled and pseudo-labeled data, and repeats, progressively refining label quality and synthesis. We validate on two expressive tasks, word-level prominence and utterance-level emotion, across multiple low-resource data splits. We find that iterative refinement can improve pseudo-label accuracy over single-pass baselines. Furthermore, we observe that these improvements in pseudo-labeling of expressivity translate to gains in expressive label adherence and synthesis quality, confirmed by objective metrics and human listening tests. In the most data-scarce conditions, ISL-trained models outperform single-pass pseudo-labeling and further approach fully supervised performance, demonstrating that gradient-based ISL is an effective solution to expressive label scarcity in low-resource TTS.
\end{abstract}


\section{Introduction}
\label{sec:introduction}

\IEEEPARstart{M}{odeling} and controlling expressive speech, such as prosodic qualities like prominence or paralinguistic qualities like emotion, remains a central challenge in text-to-speech (TTS). To achieve expressive control, TTS models must be conditioned on representations that capture meaningful variation in speaking style and provide a mechanism for controlling generation at inference time. Existing approaches generally fall into two paradigms: implicit representations learned from reference speech or latent style embeddings, and explicit representations that use symbolic conditioning signals such as emotion categories or prominence labels \cite{wang2018style, skerry2018towards, habib2019semi, lorenzo2018investigating, diatlova2023emospeech}. Recent advances in large-scale pretraining of  generative speech models have increased the popularity of implicit control methods, including reference-based conditioning, learned style representations, and natural-language prompting \cite{wang2023neural, du2024cosyvoice, zhang2026bm}. These approaches offer flexible and open-ended control, but their behavior can be difficult to predict when expressive attributes are specified through abstract text-based descriptions. Furthermore, disentangling linguistic content, speaker identity, and expressive characteristics remains an open challenge in controllable TTS, as expressive information is often encoded through acoustic correlates that simultaneously convey linguistic, paralinguistic, and speaker-specific information \cite{Scherer2003,Gobl2003, li2023styletts, zhang2026bm}. Explicit representations therefore remain promising because they provide interpretable and targeted control over perceptually meaningful expressive attributes. However, their use is fundamentally constrained by the scarcity of large-scale speech corpora annotated with expressive labels.

Acquiring expressive labels at scale remains challenging because expressive categories such as emotion and prominence require human perceptual judgments and are often expensive to annotate consistently. Existing approaches commonly address this problem by training dedicated classifiers or adapting self-supervised speech representations to automatically predict expressive labels \cite{yang2021superb,lilayer2022,de2024emphassess,ren2024emo}. While effective, these approaches introduce an additional modeling stage whose design is separate from the TTS system itself. In practice, deploying such methods requires selecting or adapting suitable classifier architectures, pretrained representations, and training procedures for each expressive task and dataset. These challenges are further compounded by variability in how expressive categories are defined, perceived, and represented across speech corpora \cite{mori2016accuracy,van2023modelling}. As a result, automatic expressive labeling often entails substantial engineering effort beyond the development of the speech synthesis model.

One effective strategy for addressing label scarcity is leveraging unlabeled data via a semi-supervised learning framework known as Iterative Self-Learning (ISL)~\cite{xie2020self,synnaeve2020endtoend}. In speech processing, ISL is frequently implemented as Iterative Pseudo-Labeling (IPL), where a model trained on limited labeled data is used to generate pseudo-labels for unlabeled examples and then retrained on the combined corpus. In this framework, the same TTS model both performs labelling and uses the labels for controllable TTS training. ISL has proven effective in low-resource speech applications, particularly for Automatic Speech Recognition (ASR), where iterative pseudo-labeling achieves competitive Word Error Rate (WER) while substantially reducing reliance on labeled training data~\cite{xu2020iterative}.

Despite its success in discriminative tasks, applying an ISL paradigm to a generative task like expressive TTS introduces a distinct challenge. In discriminative settings, label quality directly affects task performance. However, in expressive TTS a model must simultaneously adhere to expressive labels and maintain overall synthesis quality. These objectives can decouple: prior work~\cite{Sanders-IC} demonstrated that a FastPitch model trained on entirely random prominence labels yielded human-listener quality mean opinion scores approaching those of a model trained with pseudo-labels, despite having no prominence control. These prior findings suggest that synthesis quality is partially driven by data quantity, while expressive label adherence depends on label quality. Such observations motivate an iterative approach in which repeated pseudo-labeling and retraining may progressively improve expressive label quality as the model evolves, analogous to how IPL improves transcription quality in ASR.

To address expressive label scarcity, this paper proposes an ISL framework utilizing Invert-Classify~\cite{Sanders-IC}, which does not require any external classifiers or pre-trained models that may struggle with cross-domain adaptation. In this framework, a TTS model is trained on a small seed of annotated data, then frozen to pseudo-label a larger corpus of unlabeled speech via Invert-Classify, and then further trained on the combined seed data and pseudo-labeled data. By iterating this cycle, the model effectively refines its own expressive speech labels. This method relies solely on the generative model's reconstruction of expressive speech to derive labels, rather than being explicitly trained to predict expressive labels as targets.

We investigate gradient-based pseudo-labeling across two forms of expressive control: word-level prominence and utterance-level emotion. Similar to previous work, we treat prominence as a binary emphasis label indicating whether a word is perceptually prominent \cite{Sanders-IC, latif-etal-2021-controlling}. These tasks differ not only in expressive category but also in control granularity with respect to time-scale, with prominence operating locally at the word level and emotion acting globally across the utterance. Together, the two tasks are also used to investigate whether iterative gradient-based pseudo-labeling generalizes across distinct segmentation levels and expressive control settings while progressively improving expressive label quality and downstream synthesis. This leads to the following research questions:

\begin{enumerate}
\item \textbf{Generalizability:} Can iterative gradient-based pseudo-labeling generalize across expressive control tasks with differing segmentation levels and expressive characteristics, such as word-level prominence and utterance-level emotion?

\item \textbf{Iterative Dynamics:} How does pseudo-label accuracy evolve during ISL in a generative setting where labels are inferred through gradient-based pseudo-labeling, and what training schedule best balances convergence and label error propagation?

\item \textbf{Label Quality and Synthesis:} Does pseudo-label accuracy on held-out data serve as a reliable indicator of expressive label adherence and perceptual expressivity across varying seed-data conditions?

\item \textbf{Performance Limit:} To what extent does iterative refinement improve upon single-pass gradient-based pseudo-labeling, and how closely can ISL-trained models approach fully supervised performance?
\end{enumerate}

\section{Related Work}
\label{sec:related}

\subsection{Iterative Self-Learning and Pseudo-Labeling}
While the concept of self-learning is well-established~\cite{scudder1965}, typically
involving a static, single-pass approach where a teacher model labels data once for a student model, its successful application to modern deep learning relies on managing the
dynamics of an iterative loop. In ASR, IPL has emerged as a standard protocol for leveraging unlabeled data~\cite{xu2020iterative,synnaeve2020endtoend}. These works have
demonstrated that a cyclical ``pseudo-label, retrain, pseudo-label, etc.'' process prevents the model from converging prematurely to the biases of the initial seed model.

Standard IPL implementations in ASR often introduce complex heuristics, such as re-initializing models at each step or labeling only a high-confidence subset of
unlabeled data based on top-percentage softmax logits~\cite{xu2020iterative,likhomanenko2021slimipl}.
However, the efficacy of IPL is not uniform across all data distributions. Recent work on incremental semi-supervised learning for multi-genre ASR demonstrates that blindly
applying IPL to heterogeneous data can lead to performance
degradation~\cite{khonglah2020incremental}. These findings underscore the necessity of mechanisms to arrest the feedback loop before noise dominates the signal.

The mechanism underlying this balance is further formalized by Wallington et al.~\cite{wallington2021learning}, building on the foundational insight that deep networks tend to learn simple patterns early in training while memorizing noise later~\cite{arpit2017closer}. This insight is particularly relevant for expressive TTS, where ``label noise'' is not merely stochastic error but often stems from the inherent subjectivity of the task itself. For instance, inter-annotator agreement varies significantly between distinct expressive traits, such as structural prominence versus paralinguistic emotion~\cite{Roy2017}. This further motivates our investigation into generalizing the ISL framework across different expressive speech traits, as the optimal stopping point and learning dynamics may fluctuate based on the subjectivity and difficulty of the specific trait being modeled.

\subsection{Semi-Supervised TTS}

While semi-supervised learning in ASR typically revolves around iterative label refinement and pseudo-label cycling, work in semi-supervised TTS has generally taken a different route: rather than improving label quality, it focuses on leveraging unpaired
speech to improve internal representation learning.

One widely followed strategy uses vector-quantized (VQ) or otherwise discretized speech representations. Tu et al.~\cite{tu2020semi} proposed a semi-supervised multi-speaker TTS framework that uses a VQ-VAE trained with a speech reconstruction task to extract discrete acoustic units (DAUs) from large-scale untranscribed speech. These units serve as an intermediate representation that enables training without relying on aligned text--speech supervision. QS-TTS~\cite{gao-qstts2025} advances this principle by distilling speech into a more compressed VQ representation via a multi-stage, multi-codebook self-supervised framework with contrastive learning, reporting improvements in intelligibility and Mean Opinion Scores in both standard single-speaker and low-resource settings.

In contrast, approaches such as StrawNet~\cite{sharma2020strawnet} address data scarcity
from the opposite direction. Rather than compensating for missing transcripts in speech data, StrawNet assumes the scarcity lies in speech for a given text corpus, and uses a teacher model to generate synthetic speech targets from text via distillation.

While discrete representation learning substantially mitigates paired speech-text data scarcity, it does not address controllability. Closest to addressing controllability, Habib et al.~\cite{habib2019semi} proposed a semi-supervised VAE-based TTS framework that uses partial supervision to encourage latent variables to correlate with expressive attributes such as affect and speaking rate. However, the learned representations remain latent variables inferred from data rather than discrete symbolic labels, placing this work closer to implicit control methods such as Global Style Tokens (GST)~\cite{wang2018style}. As such, it does not address the scarcity of discrete expressive conditioning labels.

In summary, while semi-supervised TTS methods leverage unlabeled speech to learn acoustic representations, none address the scarcity of discrete symbolic expressive labels used for explicit controllability. Meanwhile, ISL techniques in discriminative tasks have demonstrated the effectiveness of iterative pseudo-label refinement. However, no prior work has integrated an explicit label recovery mechanism into an iterative generative learning loop for expressive speech.

In prior work~\cite{Sanders-IC}, Invert-Classify was introduced as a mechanism for recovering explicit expressive labels using a pre-trained generative model, limited to a single-pass pseudo-labeling procedure. Here, we extend this paradigm by embedding
Invert-Classify into an Iterative Self-Learning loop, treating the generative TTS model as an evolving teacher that progressively refines both synthesis and labeling quality.

\section{Methods}
\label{sec:methods}

\subsection{Iterative Self-Learning (ISL) Framework}
\label{sec:isl-framework}

We propose an ISL framework designed to progressively refine both the generative model
and the training labels. Unlike distillation approaches that rely on distinct teacher and
student architectures, this framework employs a single evolving model that learns from
its own predictions. Our methodology adopts the IPL protocols used in
ASR~\cite{xu2020iterative}. The pipeline operates as a cyclical interaction where the
model evolves through successive rounds of self-labeling and re-training.

\begin{figure}[!t]
  \centering
  \includegraphics[width=\columnwidth]{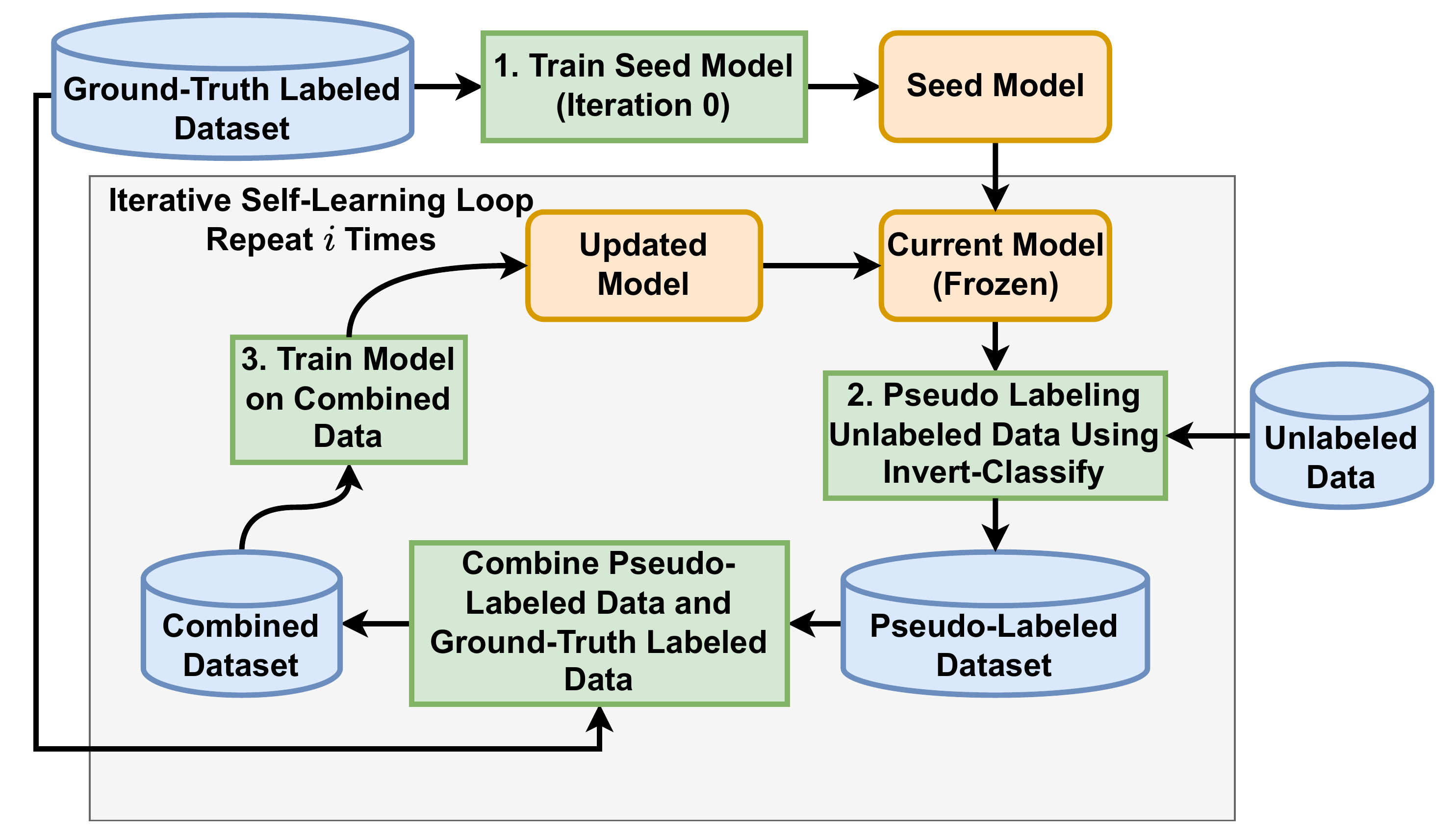}
  \caption{Overview of the Iterative Self-Learning loop.}
  \label{fig:ISL-overview}
\end{figure}

An overview of the process is shown in Fig.~\ref{fig:ISL-overview} and proceeds as
follows:
\begin{enumerate}
  \item \textbf{Initialization (Iteration 0):} A seed model $M_0$ is trained for
        $E_{\mathrm{seed}}$ epochs exclusively on the small ground-truth labeled dataset
        $\mathcal{D}_L = \{(X_L, Y_L)\}$. Following the protocol in~\cite{xu2020iterative},
        we set $E_{\mathrm{seed}} = 100$ epochs to ensure the model learns sufficient
        acoustic representations before exposure to unlabeled data.
  \item \textbf{Pseudo-Labeling:} The current model $M_i$ is frozen. Invert-Classify~\cite{Sanders-IC}, a classifier-free pseudo-labeling method described in Subsection~\ref{sec:invert-classify}, is
        applied to the unlabeled dataset $\mathcal{D}_U = \{X_U\}$, generating
        pseudo-labels $\hat{Y}$ and a pseudo-labeled dataset
        $\mathcal{D}_{PL} = \{(X_U, \hat{Y})\}$.
  \item \textbf{Iterative Training:} The model is further trained on the combined dataset
        $\mathcal{D}_{\mathrm{comb}} = \mathcal{D}_L \cup \mathcal{D}_{PL}$.
  \item \textbf{Loop:} The updated model $M_{i+1}$ is frozen for the next pseudo-labeling
        step, and steps 2--3 are repeated for $I$ iterations.
\end{enumerate}

\subsection{Matcha-TTS Backbone}
\label{sec:matcha}

To enable spectrogram generative modeling with iterative self-learning and gradient-based pseudo-labeling, we utilize Matcha-TTS~\cite{mehta2024matcha} as our backbone acoustic model. We select Matcha-TTS for several practical and methodological reasons. Recent expressive TTS systems increasingly rely on large pretrained components or speech/language-model backbones, as seen in systems such as StyleTTS2~\cite{li2023styletts}, XTTS~\cite{casanova2024xtts}, and more recent language-model-based approaches including MaskGCT~\cite{wang2025maskgct} and IndexTTS2~\cite{zhou2026indextts2}. While highly capable, these systems introduce additional pretrained dependencies and training complexity that complicate controlled iterative self-learning studies. In contrast, Matcha-TTS is trained from scratch as an acoustic model using only available speech-text data, providing a lightweight and reproducible framework for repeated self-learning experiments. Furthermore, Matcha-TTS learns alignments internally via Monotonic Alignment Search (MAS)~\cite{kim2020glow}, avoiding reliance on external forced-alignment pipelines that could otherwise introduce additional engineering dependencies. Its flow-matching formulation further places it within a broader family of successful diffusion and flow-matching based TTS models using related generative objectives~\cite{popov2021grad, chen2025f5}.

While earlier diffusion probabilistic models (e.g., Grad-TTS~\cite{popov2021grad})
formulate this denoising process using stochastic differential equations (SDEs) and
typically require many inference steps, Matcha-TTS improves inference efficiency by
adopting Optimal Transport Conditional Flow Matching
(OT-CFM)~\cite{lipmanflow}. Rather than modeling a stochastic diffusion process,
OT-CFM learns an ordinary differential equation (ODE) defining a transport trajectory
from noise to data that is close to being a straight-line path between the source and target
distributions, enabling diffusion-like quality in substantially fewer inference steps.

For text processing, we use the same setup as the original released code, \texttt{phonemizer}~\cite{phonemizer} with an \texttt{espeak-ng}~\cite{espeakng} frontend for grapheme-to-phoneme conversion.

\subsection{Modeling Expressive Representations}
\label{sec:expressive-reps}
To investigate the generalizability of our framework, we modify the Matcha-TTS acoustic model to condition on two distinct expressive representations spanning different scales of control. For \textbf{word-level prominence}, following Latif et al.~\cite{latif-etal-2021-controlling}, binary prominence tags \texttt{\textless PROM\textgreater} and \texttt{\textless NOTPROM\textgreater} are interleaved with phoneme tokens at word boundaries and learned as trainable embeddings in a separate embedding table, decoupling phonetic and prominence representations. For \textbf{utterance-level emotion}, a learned embedding corresponding to the target emotion class is concatenated to the text encoder outputs, a standard approach in emotional TTS~\cite{wang2018style,lorenzo2018investigating,diatlova2023emospeech}.

\subsection{Generative Modeling via OT-CFM}
\label{sec:otcfm}
Matcha-TTS~\cite{mehta2024matcha} is trained via Optimal Transport Conditional Flow Matching (OT-CFM)~\cite{lipmanflow}, which learns a vector field defining a trajectory from a Gaussian source to the mel-spectrogram data distribution along nearly straight lines.  While the OT-CFM objective models transport from noise to data, the auxiliary losses $\mathcal{L}_{\mathrm{dur}}$ and $\mathcal{L}_{\mathrm{enc}}$ stabilize duration predictions and encoder representations during training, using hyperparameters $\lambda_{\mathrm{dur}}$ and $\lambda_{\mathrm{enc}}$ as weights for balancing the combined fusion loss. The total training objective combines the OT-CFM loss with auxiliary duration and encoder prior losses:
\begin{equation}
\mathcal{L}_{\mathrm{Total}}
= \mathcal{L}_{\mathrm{OT-CFM}}
+ \lambda_{\mathrm{dur}}\,\mathcal{L}_{\mathrm{dur}}
+ \lambda_{\mathrm{enc}}\,\mathcal{L}_{\mathrm{enc}}
\label{eq:total_loss}
\end{equation}
The combined objective is used both for training and as the inversion loss in Section~\ref{sec:invert-classify}.
\subsection{Gradient-Based Pseudo-Labeling}
\label{sec:invert-classify}

\textbf{Invert-Classify}~\cite{Sanders-IC} is a classifier-free pseudo-labeling method for generative models. A model trained on seed labeled data is frozen, and unlabeled samples are pseudo-labeled via a two-step process. First, the missing label representation is initialized as the mean of all possible label embeddings and updated via gradient descent with respect to the same training loss used for training the seed model i.e., updated via gradient descent with respect to $\mathcal{L}_{\mathrm{Total}}$ (Equation~\ref{eq:total_loss}), a form of inversion~\cite{KINDERMANN1990277, gal2023an}. The label representation is updated toward the conditioning signal most consistent with the observed speech. Second, the updated continuous embedding is quantized by assigning it to the nearest entry in the label embedding table via cosine similarity, producing a discrete pseudo-label suitable for training.
The Invert-Classify approach bears a conceptual resemblance to classifier-free guidance~\cite{ho2021classifier} in that both leverage the generative model itself as a signal about the conditioning space, but differs in mechanism and purpose: CFG manipulates conditioning at inference time to improve sample quality, whereas Invert-Classify optimizes conditioning at the input level to recover missing discrete labels for pseudo-labeling.

\textbf{Stochastic Sampling and Inversion.}
Standard flow-matching training relies on stochastic sampling of a random timestep
$t \sim \mathcal{U}[0,1]$ and noise $x_0 \sim \mathcal{N}(0,I)$ at every batch. This
stochasticity renders inversion intractable: if $t$ and $x_0$ are resampled at every
update step of the inversion loop, the objective shifts constantly, preventing input
embeddings from converging to a stable representation.

\textbf{Adapting Inversion to Flow-Matching.}
During inversion, we adopt a deterministic environment by fixing the timestep to a
constant value (e.g., $t = 0.9$) and using a single fixed noise tensor $x_0$ across all
inversion steps for a given input. This ensures that $\nabla\mathcal{L}_{\mathrm{Total}}$
primarily reflects the conditioning labels rather than stochastic sampling variance. We treat $t$ as a hyperparameter and conduct all experiments at $t = 0.9$ (i.e., timesteps near the end of interpolation), which yielded greater gradient magnitudes with respect to the conditioning labels for the given tasks and datasets used.

\subsection{Dataset Selection and Retraining Strategy}
\label{sec:select-retrain}

Iterative pseudo-labeling introduces two practical considerations. First, pseudo-label
accuracy determines the model's ability to realize target expressive traits. Second, the
number of training epochs per iteration affects convergence and generalization. Too few
epochs may yield suboptimal synthesis, while past work has shown that running an ISL loop indefinitely risks
reinforcing errors in the pseudo-labels \cite{wallington2021learning}.

To address this, we use a \textbf{Select and Retrain} strategy. We monitor
pseudo-label accuracy on a held-out validation set across all ISL iterations, identify
the iteration $i^*$ achieving the highest validation accuracy, and train a new model from
scratch on the fixed dataset $\mathcal{D}_{\mathrm{comb}}^{(i^*)}$, decoupling the
benefits of improved labels from the model state accumulated during the iterative process.

\section{Experimental Setup}
\label{sec:experiments}

\subsection{Datasets}

\textbf{Emotion -- ESD.} We use the English subset of the Emotional Speech Dataset (ESD)~\cite{ZHOU20221_esd}, comprising 10 native English speakers and approximately 14.5 hours of speech across five emotion classes: \textit{Neutral}, \textit{Happy}, \textit{Angry}, \textit{Sad}, and \textit{Surprise}. Its parallel structure, where identical lexical content is recorded across all emotion categories, minimizes lexical bias. We follow the official train-test splits and resample audio from 16 kHz to 22.05 kHz for compatibility with HiFi-GAN.

\textbf{Prominence -- Naver-Prosody.} We use the Naver Prosody-Control dataset~\cite{latif-etal-2021-controlling}, containing 26.7 hours of speech ($\approx$36{,}600 utterances) from a single female American English speaker. Utterances are organized in contrastive focus groups with prominence placed on different constituents, from which we derive binary prominence targets (\texttt{\textless PROM\textgreater} vs.\ \texttt{\textless NOTPROM\textgreater}). Interrogative utterances are excluded, leaving groups of four parallel utterances per text. An example contrastive focus group is shown in Table~\ref{tab:data-sample}.

\begin{table}[t]
\centering
\caption{Example contrastive focus group from Naver-Prosody.}
\label{tab:data-sample}
\begin{tabular}{l}
\toprule
\textbf{Prominence-labeled utterance} \\
\midrule
\texttt{\textless NOTPROM\textgreater} Abby \texttt{\textless NOTPROM\textgreater} went \texttt{\textless NOTPROM\textgreater} home \\
\texttt{\textless PROM\textgreater} Abby \texttt{\textless NOTPROM\textgreater} went \texttt{\textless NOTPROM\textgreater} home \\
\texttt{\textless NOTPROM\textgreater} Abby \texttt{\textless PROM\textgreater} went \texttt{\textless NOTPROM\textgreater} home \\
\texttt{\textless NOTPROM\textgreater} Abby \texttt{\textless NOTPROM\textgreater} went \texttt{\textless PROM\textgreater} home \\
\bottomrule
\end{tabular}
\end{table}

\subsection{Low-Resource Data Partitioning}
\label{sec:data-partitioning}

\textbf{Stratified Sampling (ESD).}
Given the multi-speaker and multi-class nature of ESD, we use stratified random sampling
by both \textit{Speaker ID} and \textit{Emotion Class}. The target percentage is applied
individually to the utterance count within each stratum, guaranteeing full coverage of
speakers and emotion categories even in extreme low-resource settings.

\textbf{Group-wise Sampling (Naver-Prosody).}
For Naver-Prosody, we sample at the lexical group level. A ``1\% split'' refers to
sampling 1\% of available lexical groups (quadruplets), ensuring that for every text
included, all four variations of prominence placement are observed, preserving the
discriminative nature of the dataset at low data volumes.

\subsection{Model Configuration and Hyperparameters}

\textbf{TTS Training.}
We use the Adam optimizer~\cite{kingma2014adma} with a learning rate of $4 \times 10^{-4}$,
no weight decay, gradient clipping at 5.0, and a batch size of 32. All seed models are
trained for 100 epochs prior to the first IPL iteration~\cite{xu2020iterative}. For the
final Select-and-Retrain phase, models are trained for a total of 1000 epochs.
During synthesis, we use the Euler ODE sampler with temperature 1.0 and 40 steps.

\textbf{Inversion.}
Emotion embeddings are updated for 200 steps with a learning rate of 0.01; prominence
embeddings for 100 steps with a learning rate of $10^{-4}$. Both use a batch size of 32.

\textbf{Vocoder.}
We train a HiFi-GAN V1 vocoder~\cite{kong2020hifi} on the combined training splits of both datasets, using an 80-bin mel-spectrogram at 22.05\,kHz for 500{,}000 steps (batch size 16, Adam optimizer, initial learning rate $2 \times 10^{-4}$, exponential decay). The vocoder is fixed throughout all experiments, providing a consistent waveform reconstruction ceiling across systems and ensuring that observed differences arise from the acoustic models and pseudo-labeling framework rather than vocoder variation.

\subsection{Evaluation Metrics}

We evaluate along three axes: pseudo-label accuracy, objective synthesis quality, and
objective and subjective expressive-label adherence.

\textbf{Pseudo-Label Accuracy.}
\begin{itemize}
  \item \textbf{Macro F1 (Emotion):} F1 score calculated independently per class and
        averaged with equal weight across all five emotion classes.
  \item \textbf{Binary F1 (Prominence):} F1 score for the \texttt{\textless PROM\textgreater} class,
        focusing on the model's precision and recall in detecting prosodic stress.
\end{itemize}

\textbf{Synthesis Quality -- TTSDS2.}
We use the improved Text-to-Speech Distribution Score
(TTSDS2)~\cite{minixhofer2024ttsds},
which measures the 2-Wasserstein distance between synthesized and reference speech across
multiple feature spaces, normalized against a discriminative noise distribution. We report
four factors: General Quality (HuBERT, WavLM), Prosody (Pitch, MPM, Speaking Rate),
Speaker Similarity (D-Vector, WeSpeaker), and Intelligibility (Whisper, Wav2Vec2). All
TTSDS scores are in a 0--100 range where higher values indicate better performance.

\textbf{Emotion Recognition.}
We use the pre-trained \textit{emotion2vec\_plus\_large} checkpoint~\cite{ma2024emotion2vec} to
classify the emotion of synthesized audio, reporting Macro F1 score between intended and predicted emotion labels. 

\textbf{Subjective Expressivity.}
We conduct A/B preference listening tests to assess perceptual expressivity. Thirty listeners were recruited via Prolific and compensated at the platform minimum rate. Each participant completed one set of 30 questions per task. For the emotion task, participants were shown a target label and sentence, for example: \textit{Target Emotion: Neutral. Sentence: The football teams give a tea party. Which sample sounds most faithful to the emotion label `Neutral'?''} For the prominence task, participants were shown the target word marked with asterisks and in all capital letters, for example: \textit{Sentence: Anthony teaches the *SENIOR*. Which sample sounds closest to the intended emphasis on the word `senior'?''} Question and choice order were randomized. System preference scores were estimated by fitting a Bradley-Terry model~\cite{bradley1952} to the collected pairwise judgments.

\section{Experiments}
\label{sec:isl-experiments}

To address the research questions in Section~\ref{sec:introduction}, we conduct two
experiments. The first investigates Generalizability and Iterative Dynamics. The second evaluates Performance Limit \& Initialization.

\subsection{Generalizability and Iterative Dynamics}

We perform the ISL loop on diverse expressive labels using the partitioning strategies
in Section~\ref{sec:data-partitioning}. To ensure consistency, we use an identical set
of splits for both datasets, chosen to simulate different data scarcity scenarios:
\begin{itemize}
  \item \textbf{ESD (Utterance-Level Emotion):} Partitions of \textbf{0.5\%},
        \textbf{1\%}, \textbf{5\%}, and \textbf{20\%} of training data. The 0.5\%
        condition contains exactly 50 samples (one example per emotion per speaker).
  \item \textbf{Naver-Prosody (Word-Level Prominence):} Identical partitions of
        \textbf{0.5\%}, \textbf{1\%}, \textbf{5\%}, and \textbf{20\%} of training
        groups.
\end{itemize}

To empirically determine the impact of training duration on label noise propagation, we
run parallel experiments varying training duration between pseudo-labeling phases:
\begin{itemize}
  \item \textbf{Conservative (3 Epochs):} Tests whether minimal gradient updates per
        iteration prevent memorization of noise, prioritizing stability over convergence.
  \item \textbf{Standard (10 Epochs):} Consistent with established IPL
        protocols~\cite{xu2020iterative}, serving as a baseline for the
        convergence--stability trade-off.
  \item \textbf{Saturation (50 Epochs):} Identifies whether the system tends toward
        model collapse or shifts optimization toward general synthesis quality at the
        expense of expressive conditioning.
\end{itemize}

For each iteration, we checkpoint the model and evaluate pseudo-label accuracy (Macro F1
for ESD, Binary F1 for Naver) on the held-out validation set.

\subsection{Performance Limit and Initialization}

This experiment isolates the quality of the pseudo-labeled data from the quantity of training iterations. As described in Section \ref{sec:select-retrain} we follow a Select-and-Retrain methodology to train fresh models from random initialization for 1000 epochs on three datasets derived from Experiment~1:
\begin{enumerate}
  \item \textbf{Optimal ISL Data:} Pseudo-labeled by the iteration $i^*$ achieving
        highest pseudo-label accuracy (high-quality, self-refined condition).
  \item \textbf{Single-Pass Control Data:} Generated by the base model (Iteration~0),
        representing baseline Invert-Classify performance.
  \item \textbf{Reference (100\% GT):} Standard supervised model trained on 100\%
        ground-truth data, serving as the performance ceiling.
\end{enumerate}
The resulting models form the basis of the objective synthesis metrics in Subsection~\ref{subsection:objective} and the subjective listening study described in Subsection~\ref{sec:subjective}.

\section{Results}
\label{sec:results}

\subsection{Impact of Training Duration}
\label{sec:res-training-duration}

\begin{figure*}[!t]
  \centering
  \subfloat[0.5\% Data Split]{%
    \includegraphics[width=0.48\textwidth]{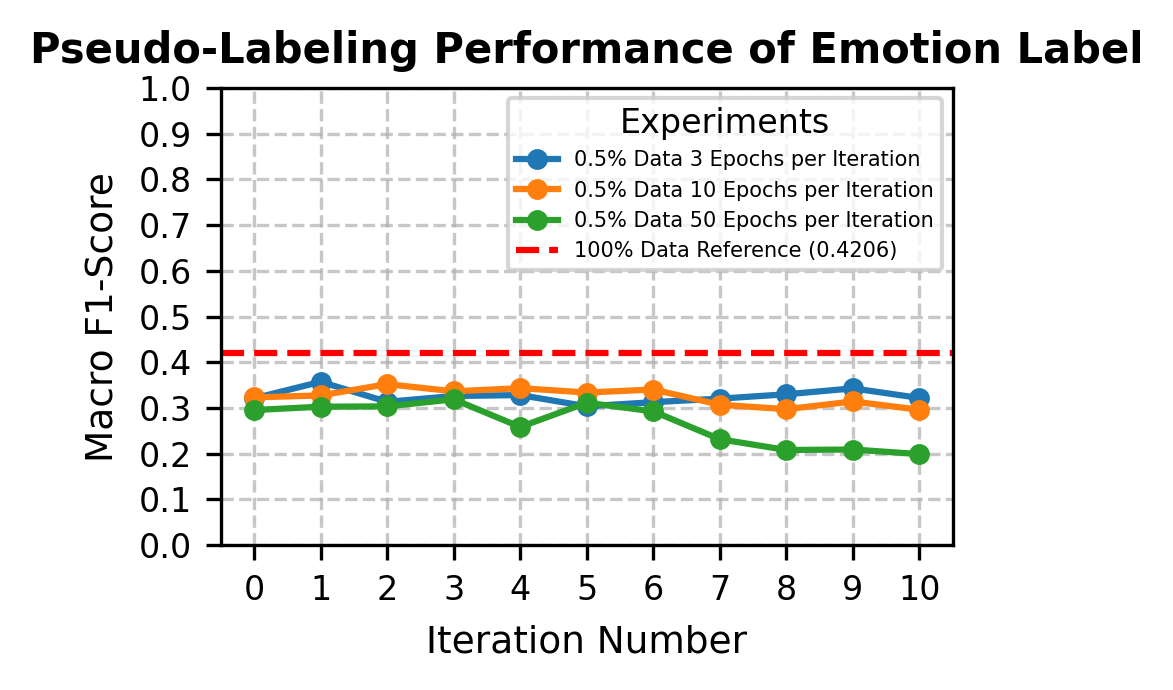}%
    \label{fig:res_a}}
  \hfil
  \subfloat[1\% Data Split]{%
    \includegraphics[width=0.48\textwidth]{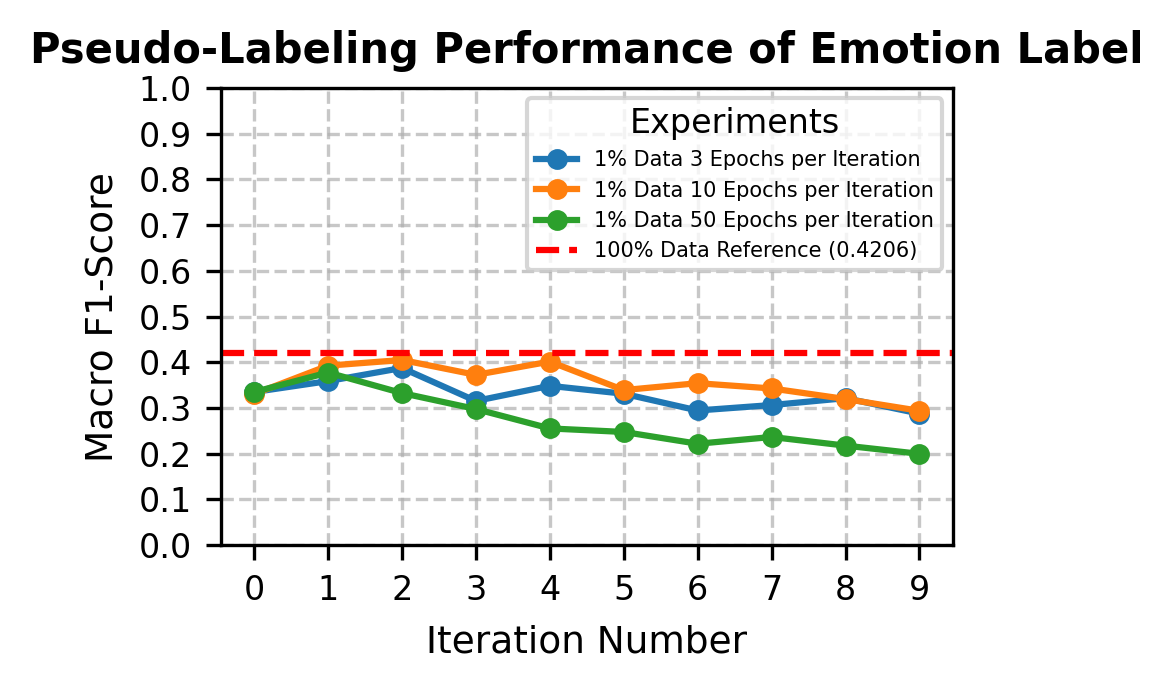}%
    \label{fig:res_b}}
  \hfil
  \subfloat[5\% Data Split]{%
    \includegraphics[width=0.48\textwidth]{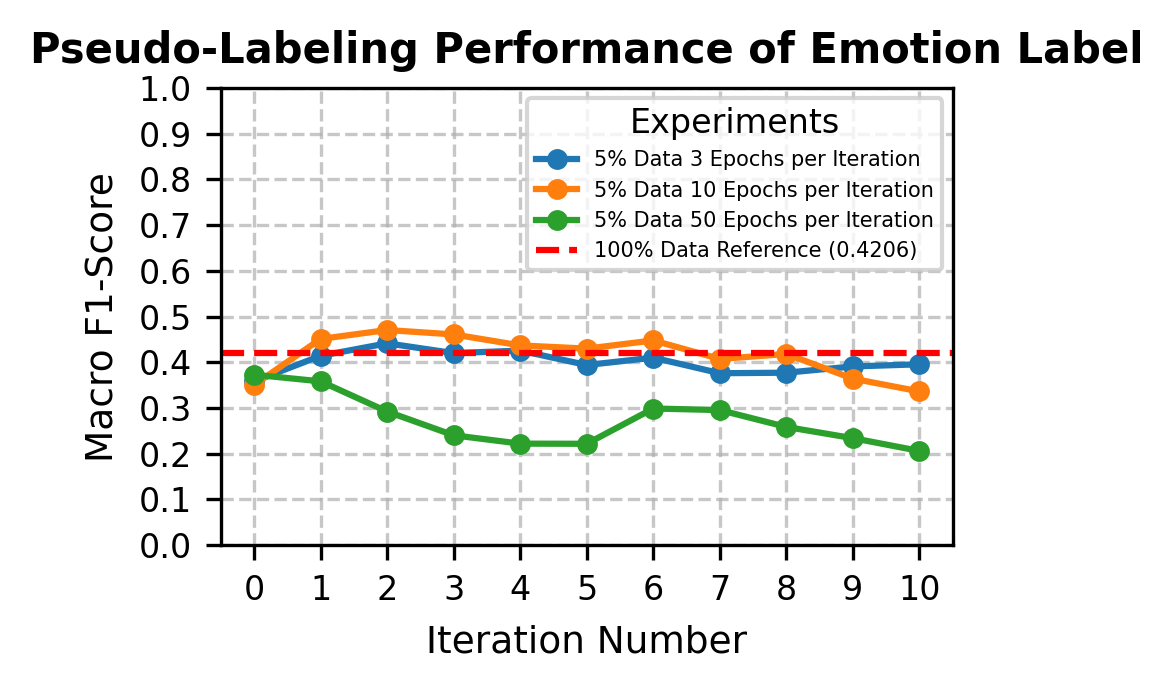}%
    \label{fig:res_c}}
  \hfil
  \subfloat[20\% Data Split]{%
    \includegraphics[width=0.48\textwidth]{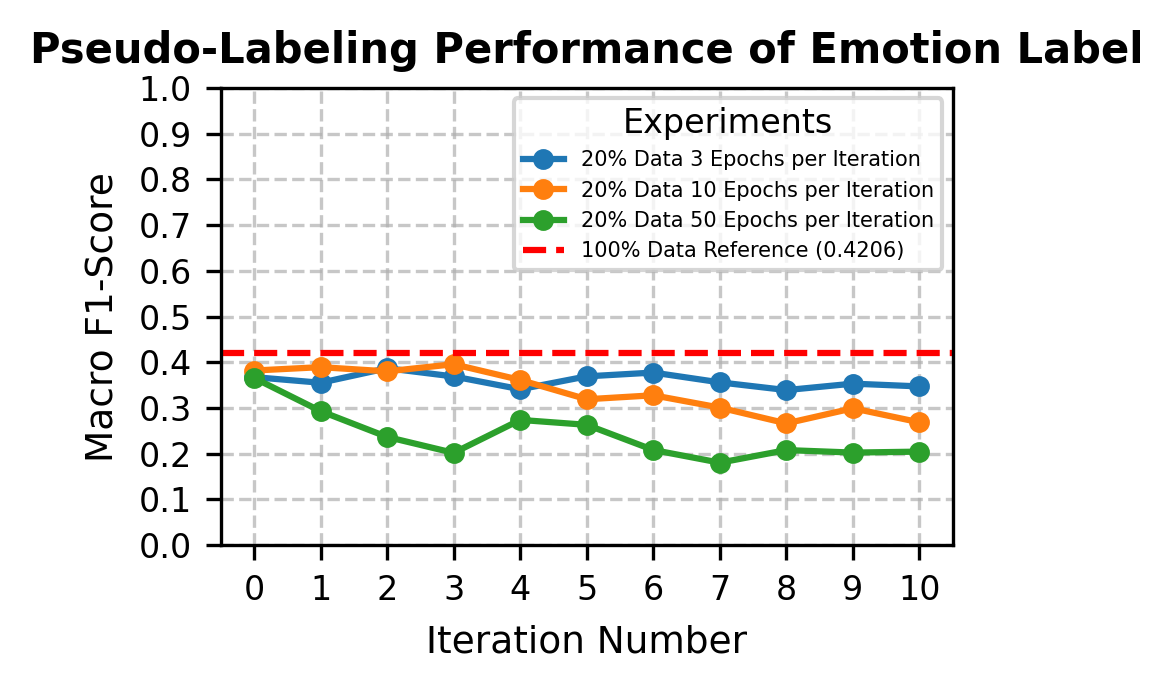}%
    \label{fig:res_d}}
  \caption{\textbf{Iterative Dynamics on ESD (Emotion):} Macro F1-Score of pseudo-labels
  on the validation set across iterations. The green line (50 epochs) illustrates model
  collapse; the orange line (10 epochs) demonstrates stable convergence.}
  \label{fig:ESD-pl-val}
\end{figure*}

\begin{figure*}[!t]
  \centering
  \subfloat[0.5\% Data Split]{%
    \includegraphics[width=0.48\textwidth]{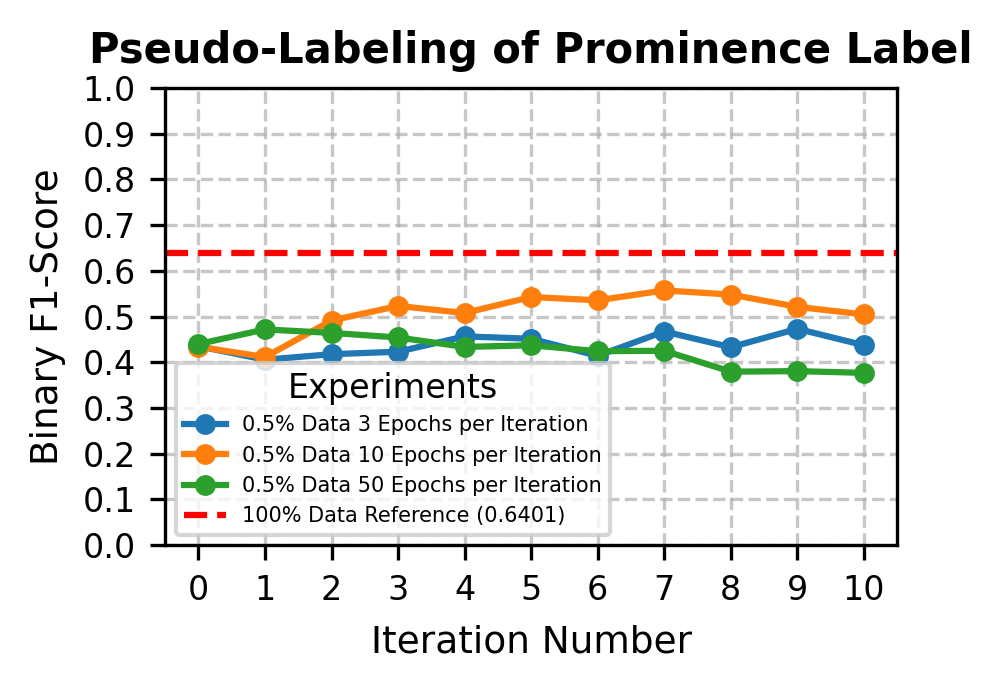}%
    \label{fig:NP_res_a}}
  \hfil
  \subfloat[1\% Data Split]{%
    \includegraphics[width=0.48\textwidth]{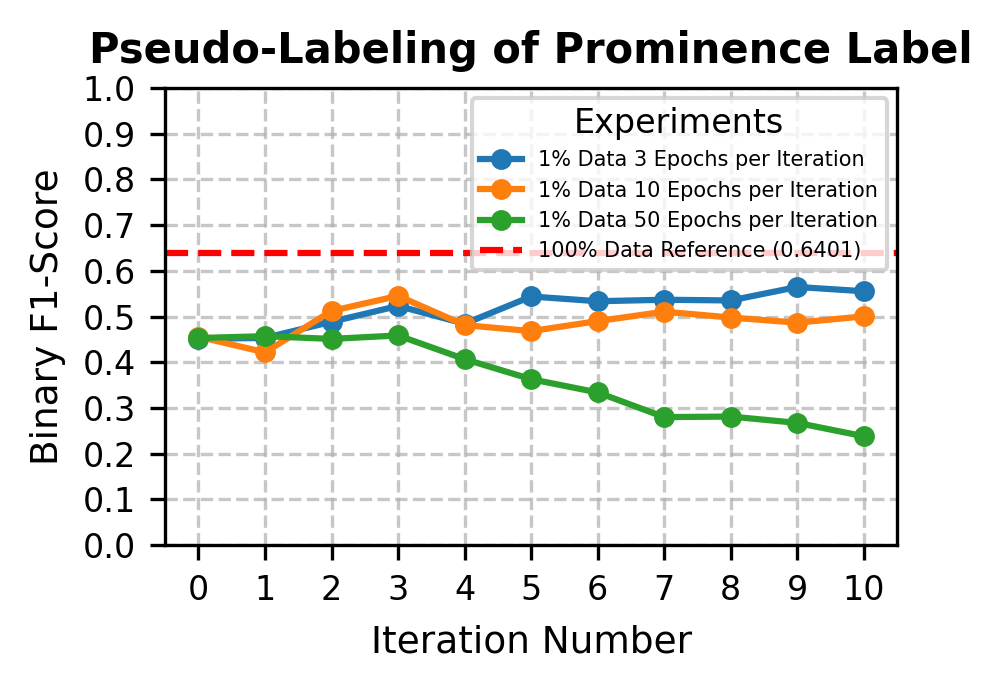}%
    \label{fig:NP_res_b}}
  \hfil
  \subfloat[5\% Data Split]{%
    \includegraphics[width=0.48\textwidth]{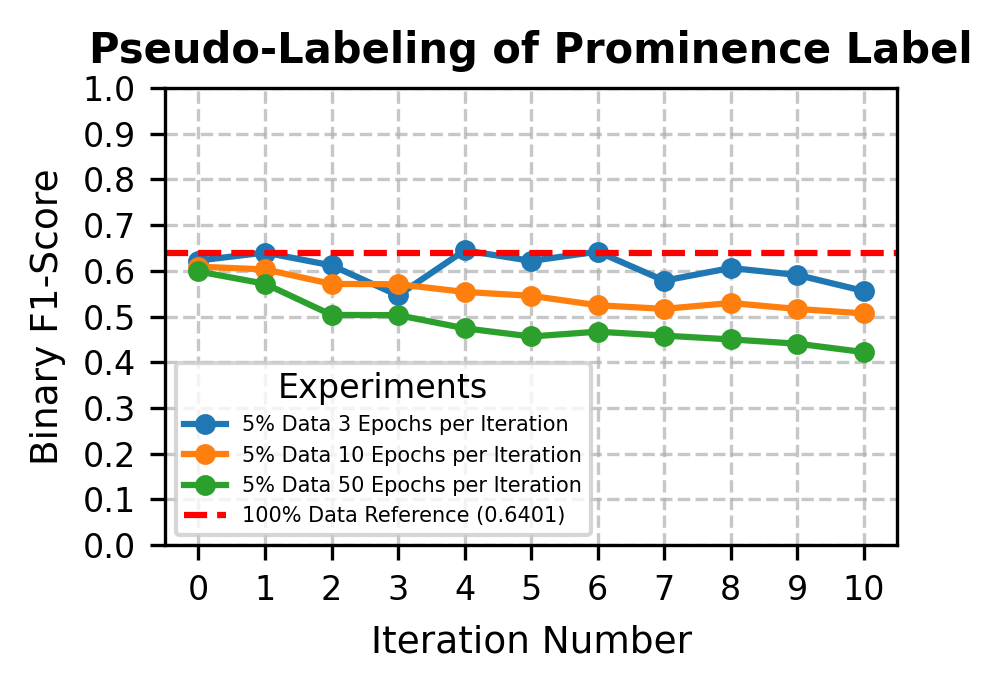}%
    \label{fig:NP_res_c}}
  \hfil
  \subfloat[20\% Data Split]{%
    \includegraphics[width=0.48\textwidth]{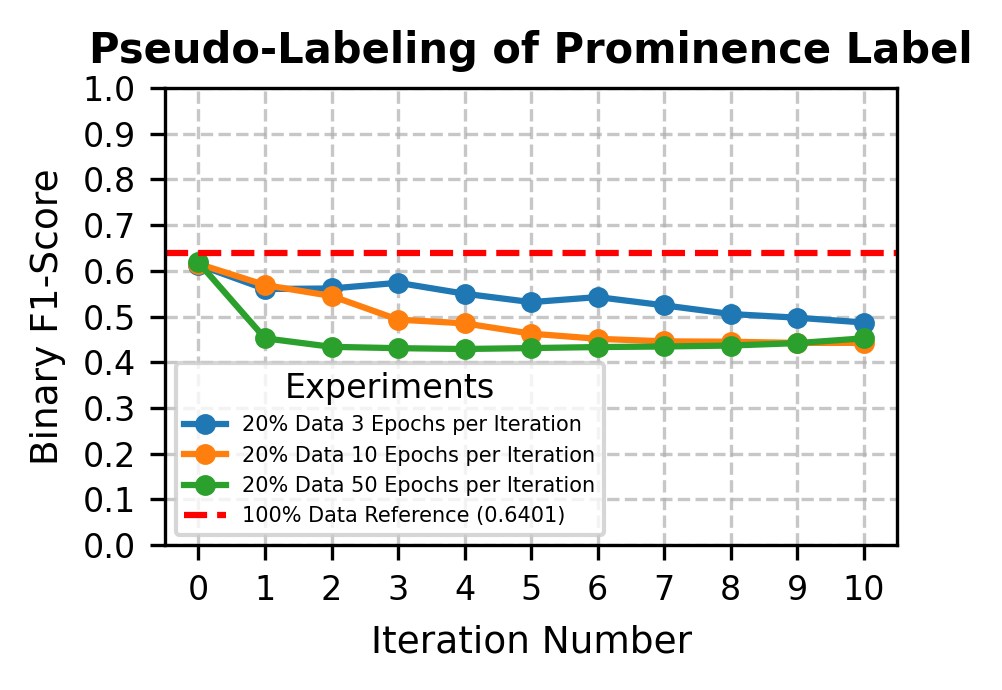}%
    \label{fig:NP_res_d}}
  \caption{\textbf{Iterative Dynamics on Naver-Prosody (Prominence):} Binary F1-Score of
  pseudo-labels on the validation set.}
  \label{fig:NP-pl-val}
\end{figure*}

The stability of the self-learning loop under varying training intensities is visualized
in Figs.~\ref{fig:ESD-pl-val} and~\ref{fig:NP-pl-val}, which show pseudo-label accuracy
on the held-out validation set across all data partitions.

A consistent failure mode is observed in the Saturation (50 Epochs) condition across both tasks: longer training between pseudo-labeling phases leads to degradation in pseudo-label accuracy. For example, in the 1\% Prominence split (Fig.~\ref{fig:NP_res_b}), the F1-score collapses from ${\approx}0.45$ to ${\approx}0.24$. All Emotion splits (Fig.~\ref{fig:ESD-pl-val}) exhibit characteristic signs of model
collapse, with pseudo-labeling accuracy degrading significantly below the seed model's initial accuracy.

In contrast, the Standard (10 Epochs) and Conservative (3 Epochs) conditions demonstrate that limiting training per iteration is crucial for stability. The 10-epoch setting
consistently yields the most robust pseudo-label accuracy for both tasks, balancing convergence speed with noise resistance. However, the 3-epoch setting occasionally achieves higher late-stage performance (e.g., Figs.~\ref{fig:NP_res_c} and~\ref{fig:NP_res_b}). We therefore identify 10 epochs per iteration as the more consistent optimal trade-off, though this remains a tunable hyperparameter with respect to a given task, dataset, and seed dataset size.

\subsection{Generalizability and Task-Dependent Learning Dynamics}

Comparing the trajectories of both datasets reveals a fundamental distinction in how ISL applies to different levels of expressive hierarchy:

For utterance-level emotion control, the primary performance gain is realized almost immediately. In the 1\% and 5\% ESD splits (the orange lines in Figs.~\ref{fig:res_b} and~\ref{fig:res_c}),
the Macro F1-Score improves most significantly within the first self-training pass (Iteration $0 \rightarrow 1$), increasing from ${\approx}0.35$ to over ${\approx}0.45$ for the 5\% split. Subsequent iterations provide marginal stability gains but saturate quickly, suggesting a high signal-to-noise ratio in the initial ground-truth labels.

Conversely, word-level prominence exhibits a slower, more gradual performance curve, with iterative refinement requiring more iterations before improvements stabilize (Figs.~\ref{fig:NP_res_a} and~\ref{fig:NP_res_b}). However, in both tasks we see either performance plateauing or degradation after reaching a local peak in pseudo-labeling accuracy over iterations, which bears some resemblance to past findings in ASR~\cite{xu2020iterative, xie2020self}. 

These differing trajectories in pseudo-labeling performance influence the number of iterations required for effective ISL and motivate the selection of different checkpoints for the final synthesis experiments reported in Subsections~\ref{subsection:objective} and~\ref{sec:subjective}.

\subsection{Impact of Data Scarcity}

The results delineate an inverse relationship between supervised data availability and the efficacy of the self-learning loop.

In the most data-scarce prominence settings (0.5\% and 1\%), ISL
yields performance significantly above the control baseline of a single iteration. For the 1\% Emotion
split, ISL improves pseudo-label F1 from ${\approx}0.33$ to ${\approx}0.40$, a relative
improvement of over 20\%. However, a lower bound to this efficacy is observed: in the
0.5\% Emotion split, pseudo-label accuracy remains flat, suggesting that a minimum
threshold of supervised signal is required to initiate bootstrapping.

As supervised data increases to 20\% (Figs.~\ref{fig:res_d} and~\ref{fig:NP_res_d}),
ISL benefits are no longer observed. This confirms that ISL functions primarily as a
low-resource intervention, yielding diminishing returns when the supervised signal is
already sufficient to learn robust generalizations.

Based on these findings, we focus subsequent synthesis analysis on the 1\% and
5\% splits for the Emotion task, and the 0.5\% and 1\%
splits for the Prominence task.
\begin{table*}[t]
\caption{\textbf{Emotion Dataset:} Comparison of Single-Pass Control vs.\ ISL at 1\%
and 5\% data splits. TTSDS-derived scores are in a 0--100 range ($\uparrow$ higher values
indicate better performance). F0 RMSE is reported in semitones ($\downarrow$ lower is
better).}
\label{tab:emotion_results}
\centering
\resizebox{\textwidth}{!}{%
\begin{tabular}{l|cc|cc|cc}
\toprule
\textbf{Metric}
  & \textbf{1\% Control} & \textbf{1\% ISL}
  & \textbf{5\% Control} & \textbf{5\% ISL}
  & \textbf{100\% GT Ref.} & \textbf{Vocoded GT} \\
\midrule
\multicolumn{7}{l}{\textit{Emotion Recognition}} \\
Emo2vec F1 Score $\uparrow$ & 26.58 & 29.87 & 39.38 & 49.58 & 45.95 & 69.32 \\
\midrule
\multicolumn{7}{l}{\textit{Prosody}} \\
F0 RMSE $\downarrow$ & 5.39 & 5.33 & 5.55 & 4.73 & 4.04 & 2.56 \\
MPM $\uparrow$         & 96.82 & 98.90 & 97.68 & 99.01 & 99.33 & 98.81 \\
Allosaurus SR $\uparrow$   & 84.46 & 89.49 & 87.79 & 89.37 & 92.86 & 95.26 \\
HuBERT Token SR $\uparrow$   & 75.81 & 76.14 & 74.58 & 74.89 & 81.71 & 86.97 \\
TTSDS Prosody Average $\uparrow$ & 85.70 & 88.18 & 86.68 & 87.76 & 91.30 & 93.68 \\
\midrule
\multicolumn{7}{l}{\textit{Speaker Similarity}} \\
D-Vector $\uparrow$  & 92.66 & 93.83 & 93.37 & 93.94 & 94.77 & 96.00 \\
WeSpeaker $\uparrow$    & 82.37 & 83.34 & 83.75 & 83.20 & 83.97 & 87.52 \\
TTSDS Speaker Average $\uparrow$ & 87.52 & 88.59 & 88.56 & 88.57 & 89.37 & 91.76 \\
\midrule
\multicolumn{7}{l}{\textit{Intelligibility}} \\
Whisper Activations $\uparrow$ & 81.92 & 81.01 & 79.68 & 81.49 & 85.27 & 84.39 \\
Wav2Vec2 Activations $\uparrow$ & 94.10 & 94.14 & 94.16 & 94.43 & 96.22 & 98.44 \\
TTSDS Intelligibility Average $\uparrow$ & 88.01 & 87.58 & 86.92 & 87.96 & 90.75 & 91.42 \\
\midrule
\multicolumn{7}{l}{\textit{General Quality}} \\
HuBERT $\uparrow$ & 94.47 & 95.28 & 94.80 & 95.36 & 95.85 & 97.32 \\
WavLM $\uparrow$ & 96.71 & 96.96 & 96.79 & 96.97 & 97.23 & 98.71 \\
TTSDS General Quality Average $\uparrow$ & 95.59 & 96.12 & 95.80 & 96.17 & 96.54 & 98.02 \\
\midrule
\multicolumn{7}{l}{\textit{Overall}} \\
TTSDS Average Scores $\uparrow$ & 88.70 & 89.90 & 89.18 & 89.85 & 92.02 & 93.71 \\
\bottomrule
\end{tabular}}
\end{table*}

\subsection{Objective Synthesis Quality and Expressive Label Adherence}
\label{subsection:objective}
\begin{table*}[t]
\caption{\textbf{Prominence Dataset:} Comparison of Single-Pass Control vs.\ ISL at
0.5\% and 1\% data splits. TTSDS-derived scores are in a 0--100 range ($\uparrow$ higher values indicate better performance). F0 RMSE (pYIN) is reported in semitones ($\downarrow$
lower is better). The TTSDS Pitch score is anomalous for this dataset due to DIO
voicing detection failure on low-F0 frames in natural speech; see
Section~\ref{sec:results} for further explanation. Absolute values should be interpreted relative to the vocoded ground-truth and 100\% reference models.}
\label{tab:prominence_results}
\centering
\resizebox{\textwidth}{!}{%
\begin{tabular}{l|cc|cc|cc}
\toprule
\textbf{Metric}
  & \textbf{0.5\% Control} & \textbf{0.5\% ISL}
  & \textbf{1\% Control}   & \textbf{1\% ISL}
  & \textbf{100\% GT Ref.} & \textbf{Vocoded GT} \\
\midrule
\multicolumn{7}{l}{\textit{Prosody}} \\
F0 RMSE $\downarrow$  & 2.66  & 2.57  & 2.68  & 2.91  & 2.55  & 2.01  \\
MPM $\uparrow$            & 97.15 & 96.87 & 96.72 & 98.65 & 97.06 & 98.51 \\
Allosaurus SR $\uparrow$   & 84.26 & 88.57 & 84.00 & 92.49 & 88.19 & 96.80 \\
HuBERT Token SR $\uparrow$ & 90.15 & 94.55 & 92.29 & 80.52 & 95.98 & 94.29 \\
TTSDS Prosody Average $\uparrow$ & 90.52 & 93.33 & 91.00 & 90.55 & 93.74 & 96.53 \\
\midrule
\multicolumn{7}{l}{\textit{Intelligibility}} \\
Whisper Activations $\uparrow$ & 84.65 & 88.95 & 86.45 & 91.33 & 87.39 & 93.41 \\
Wav2Vec2 Activations $\uparrow$ & 96.73 & 97.36 & 96.43 & 97.89 & 97.50 & 98.64 \\
TTSDS Intelligibility Average $\uparrow$ & 90.69 & 93.16 & 91.44 & 94.61 & 92.45 & 96.03 \\
\midrule
\multicolumn{7}{l}{\textit{General Quality}} \\
HuBERT $\uparrow$ & 92.75 & 93.28 & 92.67 & 92.94 & 92.94 & 96.33 \\
WavLM $\uparrow$ & 93.93 & 93.46 & 93.45 & 94.78 & 93.57 & 97.73 \\
TTSDS General Quality Average $\uparrow$ & 93.34 & 93.37 & 93.06 & 93.86 & 93.26 & 97.03 \\
\midrule
\multicolumn{7}{l}{\textit{Overall}} \\
TTSDS Average Scores $\uparrow$ & 91.37 & 93.15 & 91.57 & 92.66 & 93.23 & 96.53 \\
\bottomrule
\end{tabular}}
\end{table*}

\subsubsection{Expressive Label Adherence}

A consistent trend is observed: models trained with ISL data generally outperform Single-Pass Control models, with a specific exception in the lowest-resource prominence condition.

Significant gains are observed in the Emotion label adherence
(Table~\ref{tab:emotion_results}). The 1\% ISL condition achieves an emo2vec F1
of 29.87 vs.\ 26.58 for the Control. The gap widens at the 5\% split (49.58 vs.\ 39.38),
supporting the conclusion that pseudo-label accuracy improvements translate to better
expressive label adherence in the final synthesis.

Notably, the 5\% ISL model (49.58) outperforms the model trained on 100\% Ground Truth (45.95) in terms of Emo2vec F1 score. However, given potential domain effects and uncertainty regarding Emo2vec performance on synthetic speech, we defer to the listening test results in Subsection \ref{sec:subjective} and other objective synthesis results for further interpretation. Overall, for the emotion task, models trained using ISL pseudo-labels outperform their single-pass counterparts in downstream expressive evaluation.

For the Prominence task, no reliable pretrained classifier was available; perceptual evaluation via the listening study therefore serves as the primary measure of expressive label adherence, with results reported in Section~\ref{sec:subjective}.

\subsubsection{Prosody, Quality, and Speaker Similarity}
Beyond expressive label adherence, ISL generally maintains or improves objective synthesis quality as measured by objective synthesis metrics.

For the Prominence task (Table~\ref{tab:prominence_results}), ISL improves overall TTSDS scores relative to single-pass controls at both low-resource conditions (91.37 vs.\ 93.15 at 0.5\%, and 91.57 vs.\ 92.66 at 1\%). Improvements are particularly evident in intelligibility-related metrics, while prosody remains broadly comparable across conditions. The TTSDS Pitch metric produced an unexpected result: Vocoded GT scored lower than all TTS systems. Upon investigation, we identified voicing detection failures in DIO, the F0 extractor used by TTSDS. Our analysis revealed that 97.7\% of natural GT utterances contain voiced regions below DIO's pitch floor (72 Hz). During vocoder reconstruction, these regions were systematically shifted upward above DIO's floor, causing the Vocoded GT pitch distribution to shift upward relative to the original GT (129.4 Hz vs.\ 135.0 Hz mean F0). Because TTS systems never produced sub-100 Hz voiced F0, they were unaffected by this artifact. Therefore, we report pYIN-based F0 RMSE~\cite{pYIN}, which confirms Vocoded GT as the expected ceiling (2.01 semitones) and indicates broadly comparable pitch reconstruction between ISL and single-pass systems. ISL with 0.5\% resulted in better TTSDS Prosody Average scores and F0 RMSE in comparison to the control, however 1\% ISL did not. Beyond objective prosody metrics, ISL overall maintains or improves synthesis quality for the prominence task, as shown in the overall, intelligibility, and general quality TTSDS scores relative to single-pass controls.

For the Emotion task (Table~\ref{tab:emotion_results}), ISL similarly improves overall TTSDS scores relative to single-pass controls (88.70 vs.\ 89.90 at 1\%, and 89.18 vs.\ 89.85 at 5\%). Gains are most pronounced in prosody-related metrics and are accompanied by improved Emo2vec classification performance, suggesting that iterative refinement can improve expressive control while maintaining or modestly improving general synthesis quality.

\subsection{Subjective Evaluation}
\label{sec:subjective}
Each row observes a specific pairwise comparison related to the following: ISL versus single-pass control, the effect of additional seed data, and the performance gap between fully supervised and vocoded ground-truth anchors.

\begin{table*}[!t]
\caption{A/B preference listening test results. Win probability for System~B and 95\%
confidence intervals are estimated via the Bradley-Terry model. A win probability above
0.5 indicates preference for System~B; values with confidence intervals that overlap with 0.5 indicate a tie.}
\label{tab:ab_listening}
\centering
\begin{tabular}{lllcc}
\toprule
\textbf{Task} & \textbf{System A} & \textbf{System B} & \textbf{Preferred} & \textbf{Win Prob.\ (B) $\pm$ CI} \\
\midrule
\multicolumn{5}{l}{\textit{Naver-Prosody (Word-Level Prominence)}} \\
NP  & 0.5\% Control  & 0.5\% ISL      & ISL            & $0.656 \pm 0.117$ \\
NP  & 1.0\% Control  & 1.0\% ISL      & ISL            & $0.678 \pm 0.078$ \\
NP  & 0.5\% ISL      & 1.0\% ISL      & Tie            & $0.458 \pm 0.140$ \\
NP  & 1.0\% ISL      & 100\% GT Ref.\ & Tie  & $0.629 \pm 0.132$ \\
NP  & 100\% GT Ref.\ & Vocoded GT     & Vocoded GT     & $0.617 \pm 0.110$ \\
\midrule
\multicolumn{5}{l}{\textit{ESD (Utterance-Level Emotion)}} \\
ESD & 1.0\% Control  & 1.0\% ISL      & Tie            & $0.500 \pm 0.094$ \\
ESD & 5.0\% Control  & 5.0\% ISL      & ISL            & $0.633 \pm 0.037$ \\
ESD & 1.0\% ISL      & 5.0\% ISL      & Tie    & $0.539 \pm 0.045$ \\
ESD & 5.0\% ISL      & 100\% GT Ref.\ & 100\% GT Ref.\ & $0.633 \pm 0.042$ \\
ESD & 100\% GT Ref.\ & Vocoded GT     & Tie    & $0.533 \pm 0.062$ \\
\bottomrule
\end{tabular}
\end{table*}

In terms of subjective preference of expressivity, ISL is shown in Table~\ref{tab:ab_listening} to be preferred over single-pass control for both tasks at the more resourced splits (NP: $0.678 \pm 0.078$; ESD 5\%: $0.633 \pm 0.037$), confirming that improvements in pseudo-labeling translate to perceptually stronger expressive synthesis. At the most data-scarce conditions (0.5\% NP, 1\% ESD), ISL trained models are preferred for prominence but yield a tie with respect to the 95\% confidence intervals overlapping with a win probability of 50\% for emotion, consistent with the smaller pseudo-label accuracy gains observed at 1\% ESD. Seed data scaling comparisons for ISL trained models are ties for both tasks, suggesting that expressivity in downstream synthesis does not scale strongly with additional seed data within the ranges tested. A gap between ISL training and fully supervised with 100\% GT training data remains for the emotion task, where the reference model trained on 100\% GT data was preferred over 5\% ISL for emotion ($0.633 \pm 0.042$), but tied with 1\% ISL for prominence ($0.629 \pm 0.132$). Anchor pairwise comparisons show that vocoded GT was preferred over the fully supervised reference model for prominence ($0.617 \pm 0.110$) and tied for emotion ($0.533 \pm 0.062$).

\section{Discussion}
\label{sec:discussion}

Pseudo-label accuracy on held-out validation data proved to be a strong indicator of successful iterative self-learning. Across both emotion and prominence tasks, improvements in pseudo-label accuracy were generally accompanied by improvements in expressive label adherence, objective synthesis metrics, and subjective A/B preferences for expressivity, suggesting that the ability to recover expressive labels is closely linked to downstream controllability. However, the magnitude of these gains depended on both the expressive task and the amount of ground-truth seed data. When the initial seed set was sufficiently large, iterative self-learning often produced diminishing returns and sometimes degradation in pseudo-label accuracy, consistent with overfitting to self-generated labels. Conversely, when the seed set was extremely small, the initial model lacked sufficient supervision to bootstrap reliable improvements. Altogether, these results suggest that ISL is most effective within a bounded low-resource regime, where enough labeled data exist to initialize meaningful pseudo-labels, but expressive label scarcity remains the primary limitation. This behavior is broadly consistent with observations from iterative pseudo-labeling in ASR, where the quality of the initial model and the propagation of pseudo-label errors strongly influence the effectiveness of self-training~\cite{xie2020self,xu2020iterative}.

Furthermore, although prominence is represented as a binary label in the present work, its acoustic realization is highly variable and can be expressed through multiple combinations of pitch, duration, intensity, and voice quality~\cite{fry1955duration,murphy2018voice}. Prior work has highlighted the substantial variability involved in prosodic annotation, including variability in acoustic cues, contextual realization, and transcriber judgments, motivating approaches such as Rapid Prosody Transcription that represent prominence using continuous perceptual scores rather than strictly categorical labels~\cite{cole2016new}. Likewise, emotional expression arises from complex and speaker-dependent interactions among multiple acoustic correlates rather than a single acoustic template~\cite{Banse1996,Scherer2003}. Consequently, differences in ISL dynamics may reflect not only the expressive category being modeled, but also the degree to which the chosen label representation captures the underlying variation in the expressive phenomenon. Future work should investigate how alternative labeling schemes, including continuous prominence and affective representations like valence, arousal, and dominance used in EmoSphere++~\cite{cho2025emosphere++}, influence iterative self-learning behavior.

\section{Limitations}
A limitation of the study is that ISL was evaluated using only a Matcha-TTS backbone. However, the underlying gradient-based pseudo-labeling mechanism has now been demonstrated across both regression-based architectures~\cite{Sanders-IC} and flow-matching generative models in the present work, suggesting that the approach is not restricted to a specific TTS paradigm. Nevertheless, recent TTS systems increasingly utilize larger architectures, including diffusion-based models and LLM-based systems~\cite{li2023styletts,wang2025maskgct,zhou2026indextts2}. Whether the self-learning dynamics observed here persist at larger model scales remains an open question.

Similarly, the present work focuses on self-contained generative pseudo-labeling and its impact on downstream control in TTS, rather than external classifier-based pipelines. Direct comparison is complicated by differences in classifier architectures, supervision requirements, and even the choice of classifier itself, which may vary substantially across expressive tasks. Establishing fair comparisons between gradient-based and classifier-based pseudo-labeling under matched supervision budgets remains an important direction for future work.



\section{Conclusion}
\label{sec:conclusion}

This paper investigated whether an Iterative Self-Learning (ISL) framework based on gradient-based pseudo-labeling can address expressive label scarcity in TTS. Building on prior work in inversion-based pseudo-labeling for word-level prominence~\cite{Sanders-IC}, we extended the approach to an iterative self-learning framework and evaluated it across two distinct expressive control tasks: word-level prominence and utterance-level emotion.

The results show that iterative refinement can improve expressive pseudo-label accuracy beyond single-pass pseudo-labeling, although these improvements depend on both the expressive task and the ISL training configuration. Across both datasets, ISL was most effective within a bounded low-resource regime. Excessive retraining between pseudo-label updates often led to degradation in pseudo-label accuracy, while initializing with too little seed data provided insufficient supervision for reliable self-learning. Furthermore, the dynamics of self-learning differed across prominence and emotion, suggesting that both the expressive representation and dataset characteristics influence iterative refinement behavior.

A consistent finding across experiments was that pseudo-label accuracy on held-out validation data served as a useful indicator of downstream expressive controllability. Improvements in pseudo-label accuracy were generally accompanied by improvements in expressive label adherence, objective synthesis metrics, and subjective listener preferences. These results suggest that, unlike discriminative tasks where label accuracy and task performance are closely coupled, evaluating iterative self-learning for expressive TTS requires considering both pseudo-label quality and downstream synthesis behavior.

Several directions remain open. Future work should investigate alternative expressive representations, including continuous prominence and affective control spaces, alongside larger generative architectures. Established self-learning techniques such as confidence-based data selection, adaptive stopping criteria, and selective re-initialization~\cite{xu2020iterative,likhomanenko2021slimipl,wallington2021learning} may further improve robustness by limiting propagation of noisy pseudo-labels. Understanding how these factors interact with expressive label representations, generative training objectives, and downstream TTS controllability remains an important direction.


\bibliographystyle{IEEEtran}
\bibliography{references}

\end{document}